%% file: main.tex
\documentclass[final,5p,times,twocolumn]{elsarticle}

\usepackage{amssymb}
\usepackage{siunitx}

\newcommand*\figref[1]{Fig.\,\ref{#1}}

\begin{document}

\begin{frontmatter}
\journal{}



\title{How Surface Roughness Affects the Interparticle Interactions at a Liquid Interface}


\author[inst1,inst3]{Airi N. Kato\fnref{cor2}}
\author[inst2,inst3]{Yujie Jiang\fnref{cor2}}
\author[inst1,inst3]{Wei Chen}
\author[inst2,inst3,inst4]{Ryohei Seto\corref{cor1}}
\author[inst1,inst3]{Tao Li\corref{cor1}}
\affiliation[inst1]{organization={Wenzhou Key Laboratory of Biophysics, Wenzhou Institute, University of Chinese Academy of Sciences},
            city={Wenzhou},
            postcode={325001}, 
            state={Zhejiang},
            country={China}}
            
\affiliation[inst2]{organization={Wenzhou Key Laboratory of Biomaterials and Engineering, Wenzhou Institute, University of Chinese Academy of Sciences},
            city={Wenzhou},
            postcode={325001}, 
            state={Zhejiang},
            country={China}}

\affiliation[inst3]{organization={Oujiang Laboratory (Zhejiang Lab for Regenerative Medicine, Vision and Brain Health)},
            addressline={Address Two}, 
            city={Wenzhou},
            postcode={325001}, 
            state={Zhejiang},
            country={China}}
\affiliation[inst4]{organization={Graduate School of Information Science, University of Hyogo},
            city={Kobe},
            postcode={650-0047}, 
            state={Hyogo},
            country={Japan}}
            
\cortext[cor1]{Corresponding authors. E-mail addresses: seto@wiucas.ac.cn (R. Seto), litao@ucas.ac.cn (T. Li).}
\fntext[cor2]{These authors contributed equally to this work.}



\begin{abstract}
{\it Hypothesis:} Colloidal particles can be trapped at a liquid interface, which reduces the energetically costly interfacial area. Once at an interface, colloids undergo various self-assemblies and structural transitions due to shape-dependent interparticle interactions. Particles with rough surfaces receive increasing attention and have been applied in material design, such as Pickering emulsions and shear-thickening materials. However, the roughness effects on the interactions at a liquid interface remain less understood.\\
{\it Experiments:} Experimentally, particles with four surface roughnesses were designed and compared via isotherm measurements upon a uniaxial compression. At each stage of the compression, micrographic observations were conducted via the Blodgett method. Numerically, the compression of monolayer was simulated by using Langevin dynamics. Rough colloids were modelled as particles with capillary attraction and tangential constraints.\\
{\it Findings:} Sufficiently rough systems exhibit a non-trivial intermediate state between a gas-like state and a close-packed jamming state. This state is understood as a gel state due to roughness-induced capillary attraction. Roughness-induced friction lowers the jamming point. Furthermore, the tangential contact force owing to surface asperities can cause a gradual off-plane collapse of the compressed monolayer.
\end{abstract}


\end{frontmatter}


\section{Introduction}
\label{sec:introduction}
The morphology of colloidal particles strongly affects the interparticle interactions. Recently, by virtue of new synthetic techniques, shape-anisotropic colloids (e.g., cylinders, ellipsoids and cuboids) have been fabricated \cite{Glotzer2007} and applied for studies on morphological effects \cite{Li2019r,Sacanna2011,Ness_2022}. By being trapped irreversibly at a liquid interface, such anisotropic particles can create various self-assemblies \cite{Loudet2005,Madivala2009,Anjali2017} and undergo structural transitions upon compression in two dimensions (2D) \cite{Loudet2005,Basavaraj2006, Bordacs2006,Lewandowski2010,Li2017,Li2019r,Anjali2019}, where the capillary interaction determined mainly by quadrupole interfacial distortions%
\footnote{In case that the gravity is negligible.
More precisely, the Bond number, the ratio of the buoyancy and surface tension $\mbox{\textit{Bo}}=(\Delta\rho g L^2)/\gamma$, is sufficiently small.
($\Delta \rho$: the density difference of particle and liquid, $g$: the gravitational acceleration,
$L$: typical length scale such as a diameter of particle)}
is believed to be essential \cite{Oettel2008,Cavallaro2011,Botto2012r,Botto2012}.

The capillary attraction can also lead to a gel state which features a percolated network with a non-zero yield stress \cite{Basavaraj2006,Madivala2009}, similar to that caused by other attractive forces \cite{zaccarelli2007colloidal,Aveyard2000,Masschaele2009,Li2011}. When getting denser, hard-core interaction further helps the colloids jam and form a monolayer with much higher rigidity \cite{Trappe2001, Ji2020}. It can be highlighted that the jamming of anisotropic colloids may form monolayers with void regions, which are distinguished from the dense films of their spherical counterparts \cite{Aveyard2000,Basavaraj2006,Bordacs2006,Li2019r}. The transition from an open gel-like structure to a rigid monolayer can be generally referred to as ``close-packed jamming,'' where both capillary interaction and hard-core interaction play important roles \cite{Botto2012r}. Hereafter it is referred to as jamming for simplicity. Note that we use the term jamming, originally used for non-attractive systems, in an extended sense for attractive systems.

Similar to anisotropic colloids, particles with a rough surface also induce distortions at a liquid interface and, therefore, capillary attractions between them \cite{Stamou2000,Hooghten2013}.
Furthermore, rough particles can feature non-zero tangential contact force, as well as friction or (surface) interlocking that remarkably restrains the relative motions of contacting particles \cite{Hsu2018,Papanikolaou2013,Hsiao2017,Singh2020,Ilhan2022}. In recent studies of dense suspension rheology (i.e., 3D bulk property), this has been known to be responsible for the shift of jamming point \cite{Lootens2005,Silbert2010,Papanikolaou2013,Singh2020} and the strength of shear thickening \cite{Hsiao2017,Hsu2018,Pradeep2021,Pradeep2022}. 
Here, we should stress that what usually has been seen in rheology is the effect of friction in non-equilibrium. The role of friction in near-equilibrium processes such as quasi-static compression has yet to be understood.
Furthermore, though rough particles have been widely used in Pickering emulsions \cite{Zanini2018,Weijgertze2020}, the interparticle interactions and their dynamics at a 2D liquid interface remain to be investigated.

 In this work, the interactions between rough silica nanocolloids are explored at the air--water interface, by combining compression experiments and simulations. Four surface roughnesses were specifically designed and compared. We find that, for sufficiently rough systems, a nontrivial intermediate state appears before the close-packed jamming, which can be characterized by the formation of a percolated network. Roughness-induced capillary attraction is proven to determine this intermediate state. Meanwhile, surface roughness can lower the close-packed jamming point $\phi_\mathrm{J}$ and decrease the monolayer's order parameter at $\phi_\mathrm{J}$. By constraining the tangential motion of particles in contact, the effect of surface roughness on $\phi_\mathrm{J}$ is also verified. Numerical simulations further confirmed that the interparticle attraction can lead to a percolating structure prior to the close-packed jamming, and the interparticle friction decreases both the jamming point $\phi_\mathrm{J}$ and the crystallization order. This study provides a deeper understanding of the complex interactions between rough colloids and also elucidates the fundamental physics behind colloidal gelation and frictional jamming. These achievements can be helpful in designing exotic soft materials based on colloids' surface topography.

\section{Experimental methods}
\subsection{Particles preparation}
\label{sec:particle}
We synthesized silica particles with different roughnesses \cite{Chenwei_rough}: smooth (SM), tiny rough (TR), medium rough (MR), and very rough (VR). All particles feature microporous structures with the extrapolated pore width less than \SI{1.3}{\nano\metre} (See also the high-resolution SEM micrographs in Fig.\,S1(a)). The synthesized particles were washed with ethanol several times after synthesis. Transmission electron microscope (FEI talos, Thermo Scientific) micrographs were then taken to inspect particle morphology.
\subsection{Isotherm measurements}
Surface pressure--area ($\Pi$--$A$) isotherms ($\Pi \equiv \gamma_0-\gamma$, where $\gamma$ and $\gamma_0$ are the surface tensions with and without particles) were measured by uniaxially compressing the monolayers after spreading particles at an air--water interface in a Langmuir trough (\SI{7.5}{\centi\metre} $\times$ \SI{32.4}{\centi\metre}, KSV NIMA, Biolin Scientific). The schematic figure of the setup is shown in Fig.\,S1(b).
Before every measurement, the trough and the barriers were cleaned by brushing with ethanol and then rinsing with distilled water. Surface pressure was measured with water subphase using a platinum Wilhelmy plate (wetted length: \SI{39.24}{\milli\metre}) aligned perpendicular to the barriers approximately. The surface pressure of the subphase is confirmed to be less than \SI{0.2}{\milli\newton/\metre} during whole compression process. 
%
The particles were dispersed in isopropyl alcohol (IPA) after rinsing several times. The concentration was set \SI{20}{\milli\gram/\micro\litre}, and \SI{307.5}{\micro\litre} of the suspension was spread drop-wise at the interface. 
We waited for $30$ minutes before the start of compression to ensure that the increase of the surface pressure due to solvent becomes nearly constant and negligible. The compression was at the rate of $\SI{16}{\mm / \minute}$ ($\SI{8}{\mm / \minute}$ from the both sides), which is quasiequilibrium. The whole compression time is $\sim 1.1\times 10^3 t_\mathrm{B}$, where $t_\mathrm{B} \equiv R^2/D$ with particle radius $R$ and diffusion coefficient $D \approx \SI{1.5 e -12}{\metre^{2}/\second}$. Because of the sample consumption caused by the spreading operation and the different size/density of the particles, the initial area fraction of particles cannot be controlled accurately. A difference in the initial area fraction shifts an isotherm in the $A$ direction but does not impact the phase behavior.
\subsection{Microscopic observations}
In order to access the structural information corresponding to the measured isotherms, scanning electron microscopy (SEM) observations are carried out. According the results of the isotherms, the monolayer at aimed surface pressure was deposited on a clean silicon substrate (Zhejiang Lijing Photoelectric Technology, \SI{1}{\milli\metre} thickness) by the Blodgett method \cite{Acharya2009,Li2019r}. The substrates quickly dried after pulling out to the air. After \SI{0.3}{\nano\metre} platinum sputtering (EM ACE600, Leica), we took SEM micrographs (SU8010, Hitachi).
\section{Experimental results and discussions}
\label{sec:results_and_discussions}

\begin{figure}[tb]
\includegraphics[width=8.5cm]{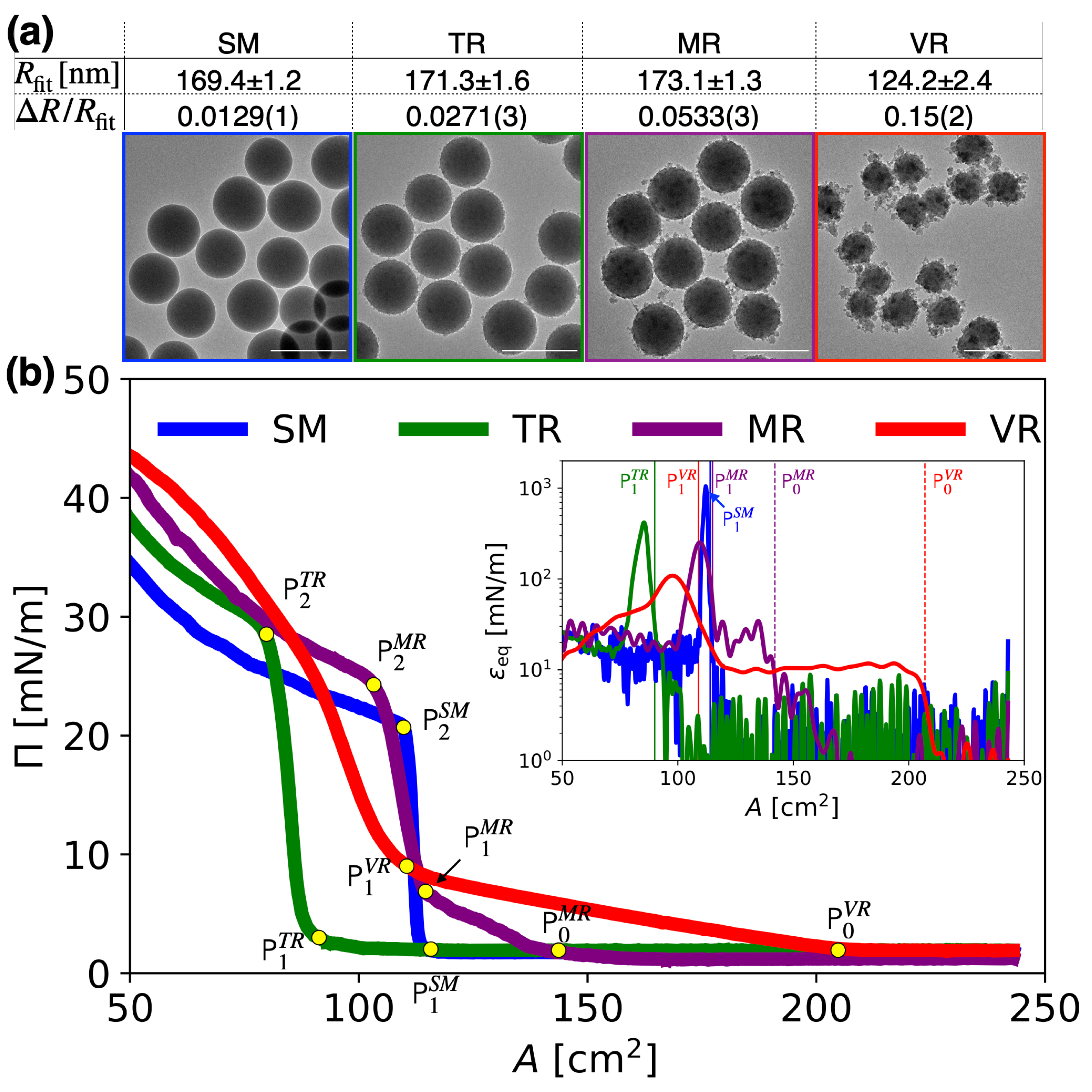}
\caption{\label{fig:1} The interfacial properties of monolayers containing SM, TR, MR, and VR particles. (a) Particle radii, roughness $\Delta R/R_\mathrm{fit}$, and the transmission electron microscope (TEM) micrographs. The scale bar shows \SI{500}{\nano\metre}. (b) $\Pi$--$A$ isotherms. The inflection points ($\mathsf{P}_{\mathrm{sequence}}^{\mathrm{roughness}}$) were defined to separate the states. The reproducibility of isotherms, except for shifts in the $A$ direction, was checked at least three times. The inset shows the dilational elastic modulus $\epsilon_\mathrm{eq}(A)$ calculated from the isotherms. (High frequency modes were removed to reduce noise.)}
\end{figure}
\subsection{Particles characterization}
TEM micrographs of SM, TR, MR, and VR particles are shown in \figref{fig:1}(a).  The particle radii and roughnesses are evaluated (\figref{fig:1}(a)) as follows. The contours of $30$ particles for each roughness are captured after binarization. The radii of particles $R_\mathrm{fit}$ are determined by the least square fitting to a circle, giving the center of the circle at the same time. We define the roughness by the standard deviation of the radial length $R(\theta)$ around the center: $\Delta R \equiv \sqrt{\langle(R(\theta)-R_\mathrm{fit})^2\rangle_\theta}$, where $\langle \cdot \rangle_\theta$ means averaging over $\theta \in [0,2\pi]$. The characterization of rough particles using the standard deviation was selected as it effectively quantifies the strength of capillary interactions, which will be examined in the subsequent section. Furthermore, the utilization of the standard deviation has been demonstrated to effectively capture the frictional nature that leads to shear thickening \cite{Hsiao2017}. It is important to mention that the fine surface structures of particles discussed in Sec.\,\ref{sec:particle} correspond to higher orders of interface deformations, which do not significantly contribute to capillary interactions.

\subsection{Isotherm measurements}
As previously observed for smooth rigid spheres, the corresponding isotherm (the blue curve, \figref{fig:1}(b)) shows a direct transition from a gas-like state (where $\Pi\sim  \SI{0}{\milli\newton/\metre}$
with a negligible interparticle interaction experimentally\footnote{According to the Gibbs adsorption isotherm, the isotherm in a gas state are $\Pi \propto 1/A$ \cite{Butt2003}. However, the $\Pi$ in a gas state is generally too small to be measured by the Langmuir trough, $\Pi <$\SI{e-4}{\milli\newton/\metre} in our case, for example.}) to the solid-like state (from $\mathsf{P}_1^{\mathrm{SM}}$ to $\mathsf{P}_2^{\mathrm{SM}}$, where particles are jamming and $\Pi$ increases sharply);
the monolayer buckles and collapses abruptly under further compression \cite{Aveyard2000,Basavaraj2006,Bordacs2006,Li2017,Li2019r} after $\mathsf{P}_2^{\mathrm{SM}}$.

%
The isotherm of TR particles (the green curve, \figref{fig:1}(b)) exhibits similar phase transitions. Isotherms for MR and VR particles (the purple and red curves, \figref{fig:1}(b)), notably, feature an intermediate state between the gas-like and solid-like states from $\mathsf{P}_0$ to $\mathsf{P}_1$ \footnote{The intermediate state is reminiscent of the liquid state of amphiphilic molecular monolayers because of the similar isotherms \cite{Kaganer1999}. However, the state should be distinct since our system consists of attractive nanocolloids (cf.~\cite{Li2017}).}.
Further compression of these monolayers induces less clear inflection points with roughness. Specifically, $\Pi$ rises more moderately after the solid-like state for the VR system (the red curve, \figref{fig:1}(b)).

Previous studies of rod-like particles have pointed out that the gradual evolution from a gas-like to solid-like state is attributed to the formation of a particle percolating network, where the lateral capillary interaction plays an essential role \cite{Basavaraj2006,Li2017}.
In the case of rough colloids, roughness-induced interfacial distortions yield capillary attractions, where the capillary interaction energy $U_\mathrm{cap}$ is proportional to the square of the liquid interfacial height undulation $\Delta H$ \cite{Stamou2000,Hooghten2013}, i.e., $U_\mathrm{cap}\propto (\Delta H)^2$. On the other hand, $\Delta H$ is considered comparable with the particle’s surface roughness \cite{Zanini2017}. Therefore, $U_\mathrm{cap}\propto (\Delta R/R_\mathrm{fit})^2$. In this case, the strong capillary interactions between our VR particles can easily trigger a gelation process, which is reflected in the observed intermediate state of the isotherm (from $\mathsf{P}_0^{\rm VR}$ to $\mathsf{P}_1^{\rm VR}$, \figref{fig:1}(b)).
The interaction energies $U_\mathrm{cap}$ for MR, TR, and SM particles are approximately \SI{12}{\percent}, \SI{3.2}{\percent}, and \SI{0.72}{\percent} of that of VR particles, respectively. As a result, the intermediate state is less apparent for MR particles and nearly invisible for TR and SM particles.

The inset of Fig.~1(b) shows the dilational (compressional) elastic moduli $\epsilon_\mathrm{eq}(A)$ \cite{Zang2010,Cicuta2003,Basavaraj2006} calculated from the isotherms by 
\begin{equation}
\epsilon_\mathrm{eq}(A) \equiv -\frac{\mathrm{d} \Pi}{\mathrm{d} (\log A)}.
\end{equation}
We remark that the moduli are intensive quantities and can also be expressed as a function of the area fraction $\phi$: $\frac{\mathrm{d} \Pi}{\mathrm{d} (\log \phi)}$.
For MR and VR particles, a slight increase of $\epsilon_\mathrm{eq}(A)$ can be noticed in the intermediate state (see the red and purple curves), corresponding to the formation of a percolating network. After that, it starts increasing significantly and reaches $\sim \SI{e2}{\milli\newton/\metre}$ in the solid-like state. Such considerable elasticities are comparable to that of a densely packed monolayer of rigid spheres (e.g., our SM particles. See the blue curve) \cite{Zang2010,Cicuta2003,Basavaraj2006}, which indicates that jamming of MR and VR particles can be achieved after the intermediate state. 

When considering the capillary interaction only, surface roughness has been theoretically predicted to enhance the elasticity of particle monolayers \cite{Lucassen1992,Danov2005}. However, our results show an apparent contradiction, where the MR and VR monolayers exhibit lower elasticities than the smoother systems (\figref{fig:1}(b)). 

This contradiction suggests that another effect, not capillary interaction, is dominant in the solid-like state. Recent rheological studies of rough particles have indicated the importance of frictional contact forces \cite{Hsu2018,Papanikolaou2013,Hsiao2017,Singh2020,Silbert2010}. They are also expected to influence structural changes (including the off-plane buckling, which seems to cause the gradual collapse of the VR curve after $\mathsf{P}_1^{\rm VR}$) and elastic properties in relatively packed regimes.

\begin{figure*}[!t]
\includegraphics[width=17.5cm]{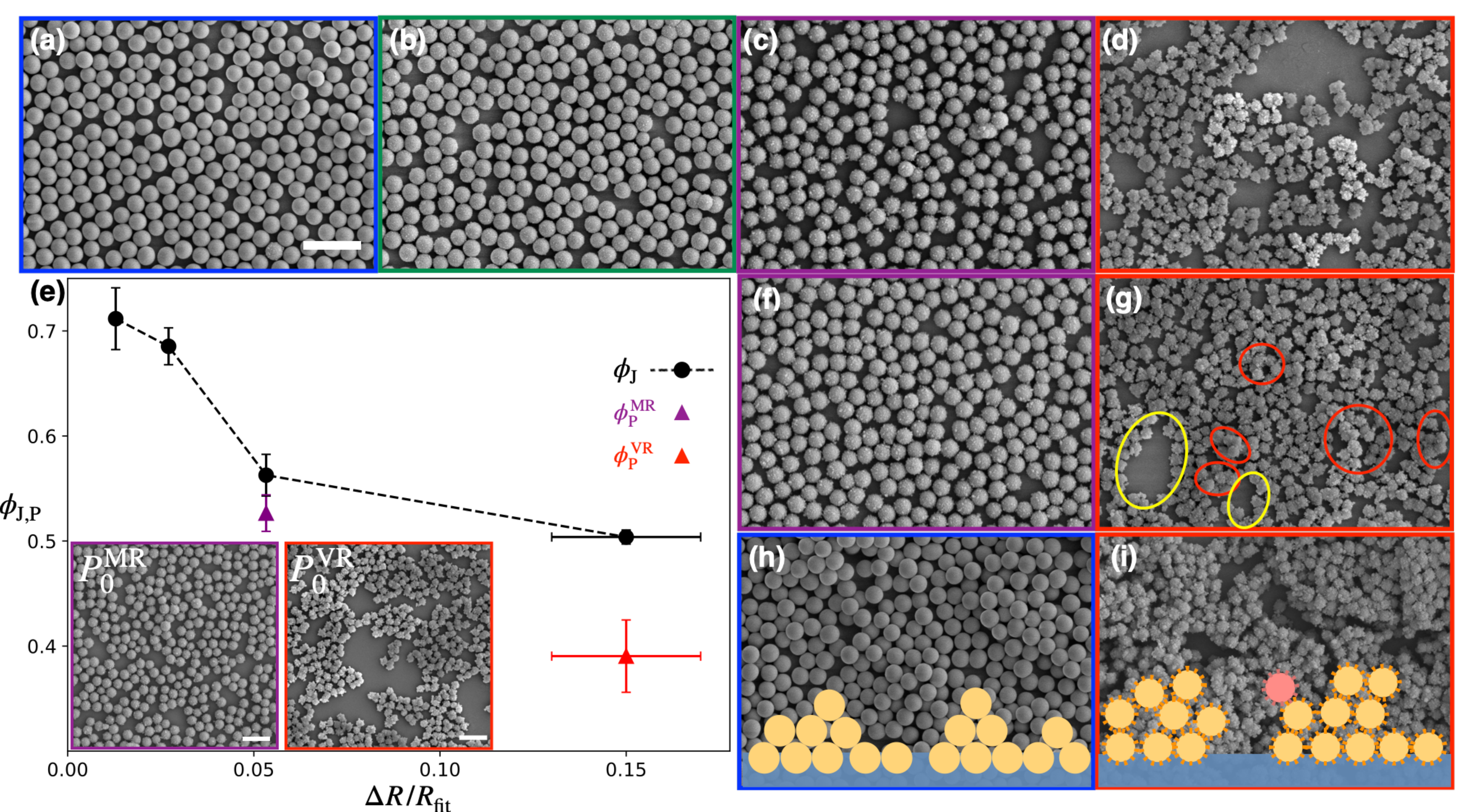}
\caption{\label{fig:2} Configurations and packing fractions at transition points.
All scale bars are \SI{1}{\micro\metre}.
(a--d) SEM micrographs at the jamming point $\mathsf{P}_1$ of (a) SM, (b) TR, (c) MR, and (d) VR. These frame colors (blue, green, purple, and red) correspond to the colors showing the different roughnesses in \figref{fig:1}. (e) The jamming point $\phi_\mathrm{J}$ and percolation point $\phi_\mathrm{P}$ were statistically analyzed as a function of roughness. More than three separate trials were performed under the same experimental conditions for each surface roughness. The insets show micrographs at the percolation point $\mathsf{P_0}$ for MR and VR. (f,g) Micrographs of the area around where collapses occurred. (f) MR at $\sim \mathsf{P}_2^{\mathrm{MR}}$, (g) VR at $\Pi\approx
\SI{15}{\milli\newton/\metre}$.
Red and yellow circles indicate noticeably ``escaped'' particles from the monolayer and voids, respectively. (h,i) Micrographs at the end of compression. (h) SM at $\Pi\approx \SI{43}{\milli\newton/\metre}$,
(i) VR at $\Pi\approx \SI{49}{\milli\newton/\metre}$.
Each schematic figure shows the expected vertical configuration. VR particles achieve an arrangement that is not stable with smooth particles by friction (e.g., see the red particle).}
\end{figure*}

\subsection{Micrography}
\label{sec:micrography}

%
Figures\,\ref{fig:2}(a)--(d) show the configurations at $\mathsf{P}_1$ of the isotherms, where all systems achieve the jamming state. Both SM and TR particles form dense monolayers (\figref{fig:2}(a),(b)).
However, voids can still be found for the MR particles (\figref{fig:2}(c)) and even more prominent for the VR particles (\figref{fig:2}(d)).
Clearly, the jamming point $\phi_{\rm J}$ decreases with the particles' surface roughness (\figref{fig:2}(e)). Combining with the previous studies of frictional granular/colloidal systems \cite{Silbert2010,Papanikolaou2013,Singh2020}, this reduction is believed to be caused by interparticle friction. As seen in \figref{fig:2}(a)--(d), more rough particles show a more disordered jammed structure, which is quantified via the global hexagonal order $\bar{\Psi}_6$ defined as follows. For $i$\,th particle, the local order parameter $\Psi_{6}(\mathbf{r}_i)$ is defined as $\Psi_{6}(\mathbf{r}_i) \equiv \frac{1}{6}\left| \sum_{j=1}^6 e^{6i\theta_{ij}}\right|$, where $\theta_{ij}$ denotes the bond angle between $i$\,th and $j$\,th particles. The global hexagonal order is then given by $\bar{\Psi}_6 \equiv \frac{1}{N}\sum_{i=1}^N \Psi_{6}(\mathbf{r}_i)$. The results are shown in Supplementary Information Sec.~II.
Furthermore, for the MR and VR systems, a percolated network is created at $\mathsf{P}_0$ (see the inset of \figref{fig:2}(e)), caused by roughness-induced capillary attraction. Indeed, the percolation point of the VR particles $\phi_{\rm P}^{\rm VR}$ is smaller than that of the MR particles $\phi_{\rm P}^{\rm MR}$, exhibiting more sparse network as a signature of stronger attractions (\figref{fig:2}(e)).


After $\mathsf{P}_1^{\rm MR}$, MR particles preferentially occupy the voids, creating a denser and more ordered monolayer than the one at $\mathsf{P}_1^{\rm MR}$ (see the change between \figref{fig:2}(c) and (f)). In the solid-like state of the VR system (\figref{fig:2}(g)), particles randomly ``escape'' from the monolayer, forming 3D aggregations at the interface (marked in red). Surprisingly, the aggregation coexists with void regions (marked in yellow). Together, these observations explain the absence of an abrupt collapse in the $\Pi$--$A$ isotherm (the red curve, \figref{fig:1}(b)). At the end of the compression, differ significantly from the SM particles that form a multi-layer with an approximately hexagonal close packing (\figref{fig:2}(h)), the VR particles create rather disordered 3D structures. As discussed above, rough particles can have considerable tangential contact force, which strongly enhances the particles’ out-of-plane escaping. The disordered structures include particles in some unstable positions (see the schematic in \figref{fig:2}(i)), which might greatly benefit from the interlocking or friction.

\section{Simulations}
\label{sec:simulation}

\subsection{Methods}

To better understand our results, we simulate the uniaxial compression of the colloidal monolayer by using the Langevin Dynamics in LAMMPS \cite{LAMMPS,islam2021normal}. We model colloids as elastic spheres (of diameter $d$ and mass $m$) with energy costs when overlapped, while a Langevin thermostat is applied to enforce thermal energy $k_{\rm B}T$. We simulate $N = 12478$ particles in a rectangular box ($700d \times 100d$) with periodic boundaries along both $x$ and $y$ directions. We prepare initial configurations by starting from a non-overlapping configuration and evolving to near equilibrium at rest. Compression is then performed by reducing the length in the $x$ direction from $L_x = 700d$ to $100d$ at a constant speed $v$ without changing $L_y$. Compared with the Brownian timescale $t_{\rm B}$, the compression is slow enough ($v \ll d/t_{\rm B}$) to ensure almost quasi-static compression as in the experiments we compare. We also consider the parameters to make the Stokes number ${\rm St} \equiv mv/d^2\eta$ sufficiently small to reduce the influence of inertia during compression.

We approach the roughness-induced interaction by considering radial attraction and tangential constraint separately. Since the capillary interaction decays fast with distance $r$ ($U_{\rm cap} \propto r^{-4}$)~\cite{Cavallaro2011,Botto2012,Botto2012r}, we assume that the Derjaguin--Muller--Toporov (DMT) model \cite{derjaguin1975effect}, a cohesive contact model, can capture the effect of the capillary interaction. The normal force is composed of elastic repulsion (with Young's moduli $E$ $\gg \eta v/d$) and surface attraction (with strength $U_{\rm att}/k_{\rm B}T$ varied from $0$ to $20$). The tangential component considers frictional resistance on sliding and rolling in a modified Coulomb manner \cite{marshall2009discrete,thornton1991interparticle}. For simplicity, we use the same spring constant $k = 10^4$ and friction coefficient $\mu$ (varied from $0$ to $1$, corresponding to frictionless and frictional, respectively) for both motions. See Supplementary Information for more details.

\begin{figure}[!t]
\includegraphics[width=8.6cm]{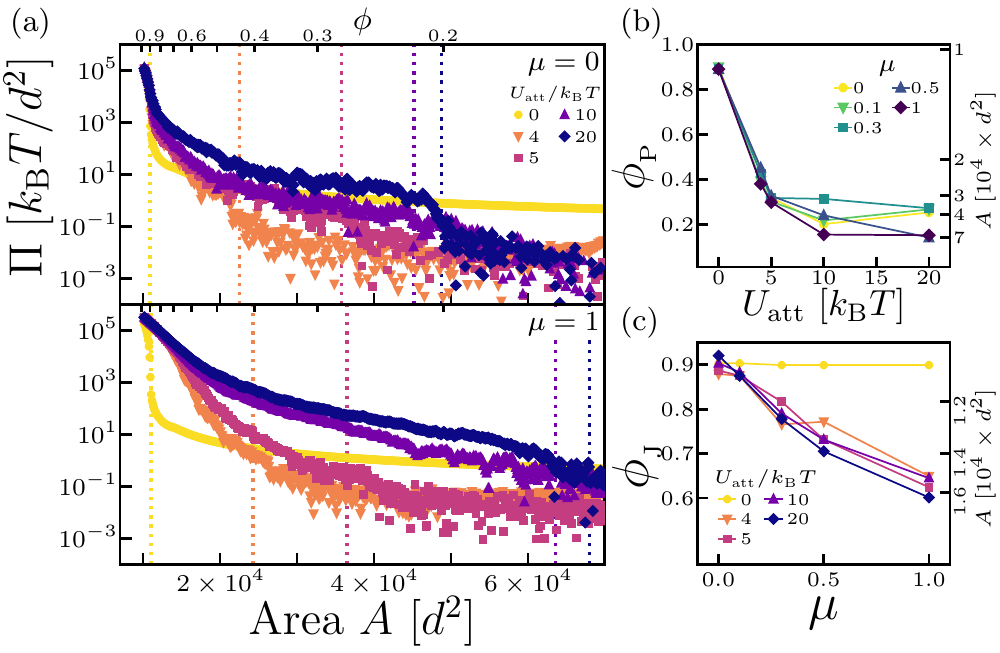}
\caption{(a) $\Pi$--$A$ isotherms of frictionless ($\mu=0$, upper) and frictional ($\mu = 1$, lower) particles with different $U_{\rm att}$. The dashed lines represent percolation points $\phi_{\rm P}$. (b) Percolation points $\phi_{\rm P}$ as functions of attraction strength $U_{\rm att}$. (c) Jamming points $\phi_{\rm J}$ as functions of friction coefficient $\mu$.}
\label{fig:3}
\end{figure}

\subsection{Results and discussions}
\label{sec:simu_results_and_discussions}

Numerical results of $\Pi$--$A$ isotherms are shown in \figref{fig:3}(a). For non-attractive particles, either frictionless ($\mu = 0$) or frictional ($\mu = 1$), the pressure $\Pi$ increases continuously as compression proceeds, consistent with theoretical calculation on hard spheres~\cite{yurkovetsky2008particle}. In the absence of attraction, frictional contacts are always relaxed at the quasi-equilibrium limit, so that friction does not contribute to isotherm at $U_{\rm att} = 0$.

Attraction with $U_{\rm att}/k_{\rm B}T = 4$, by contrast, produces an intermediate regime for both $\mu = 0$ and $\mu = 1$ cases. Due to the condensing nature, attraction reduces the pressure $\Pi$ in the gas-like state for all $U_{\rm att} > 0$ and drives colloids to form clusters that percolate the system at an early stage of compression. This results in the intermediate state in the $\Pi$--$A$ isotherm, which finally collapses on the same curve with non-attractive particles upon jamming. In the presence of attraction, we notice that the interparticle friction makes a difference in the isotherm by inducing percolation at earlier points.

We therefore attribute the intermediate state to the loose percolating at an early stage of compression. Since our compression is nearly quasi-static, colloids have sufficient time to diffuse and aggregate into a ramified network, which percolates the box at low concentrations. The intermediate state of the isotherm thereby corresponds to the compaction of colloidal gels.

Through structure analysis, we define the percolation point $\phi_{\rm P}$ as the moment when there exists at least one cluster connecting through the periodic boundary in the $x$ direction.
The dependence of $\phi_{\rm P}$ is shown in \figref{fig:3}(b). While non-attractive particles percolate at $\phi_{\rm P} \approx 0.9$, consistent with 2D random close packing \cite{Meyer_2010, zaccone2022explicit}, the attraction of $U_{\rm att} = 20k_{\rm B}T$ enables percolation at a much lower coverage $\phi_{\rm P} \approx 0.2$.
At relatively weak $U_{\rm att} \lesssim 5k_{\rm B}T$, the friction coefficient $\mu$ appears to play a minor role in both the emergence of the intermediate state (\figref{fig:3}(a)) and the percolation point $\phi_{\rm P}$ (\figref{fig:3}(b)). By contrast, strong attraction with friction leads to percolation at a lower concentration, which in turn increases the pressure of the intermediate state at the early stage of compression, \figref{fig:3}(a) and (b). These results suggest that the intermediate state (from $\mathsf{P}_0$ to $\mathsf{P}_1$) in our experiments results from the percolation transition and is highly related to the capillary attraction.

In comparison, the jamming point $\phi_{\mathrm{J}}$, corresponding to $\mathsf{P}_1$ in experiments and here defined by the sudden increase in particle overlaps (see Supplementary Information), does depend on interparticle friction. In particular, $\phi_{\mathrm{J}}$ decreases from about $0.9$ to $0.6$ as $\mu$ increases up to $1$. This agrees with our observations in \figref{fig:2}(a)--2(d), where the MR and VR particles show lower packing fractions than the smoother ones at $\mathsf{P}_1$.

During the compression in the intermediate regime, the stress relaxation is implemented through particle rearrangement so that a gel transforms into a denser gel without squeezing particles. Beyond jamming, however, elastic and plastic tangential displacements are no longer available, so the overlap of particles is required for further compaction. This scenario explains the rapid increase in $\Pi$ beyond $\mathsf{P}_1$, because the elasticity of particles is much higher than that of gels; see the inset of \figref{fig:1}(b).

\begin{figure*}
\includegraphics[width=17.5cm]{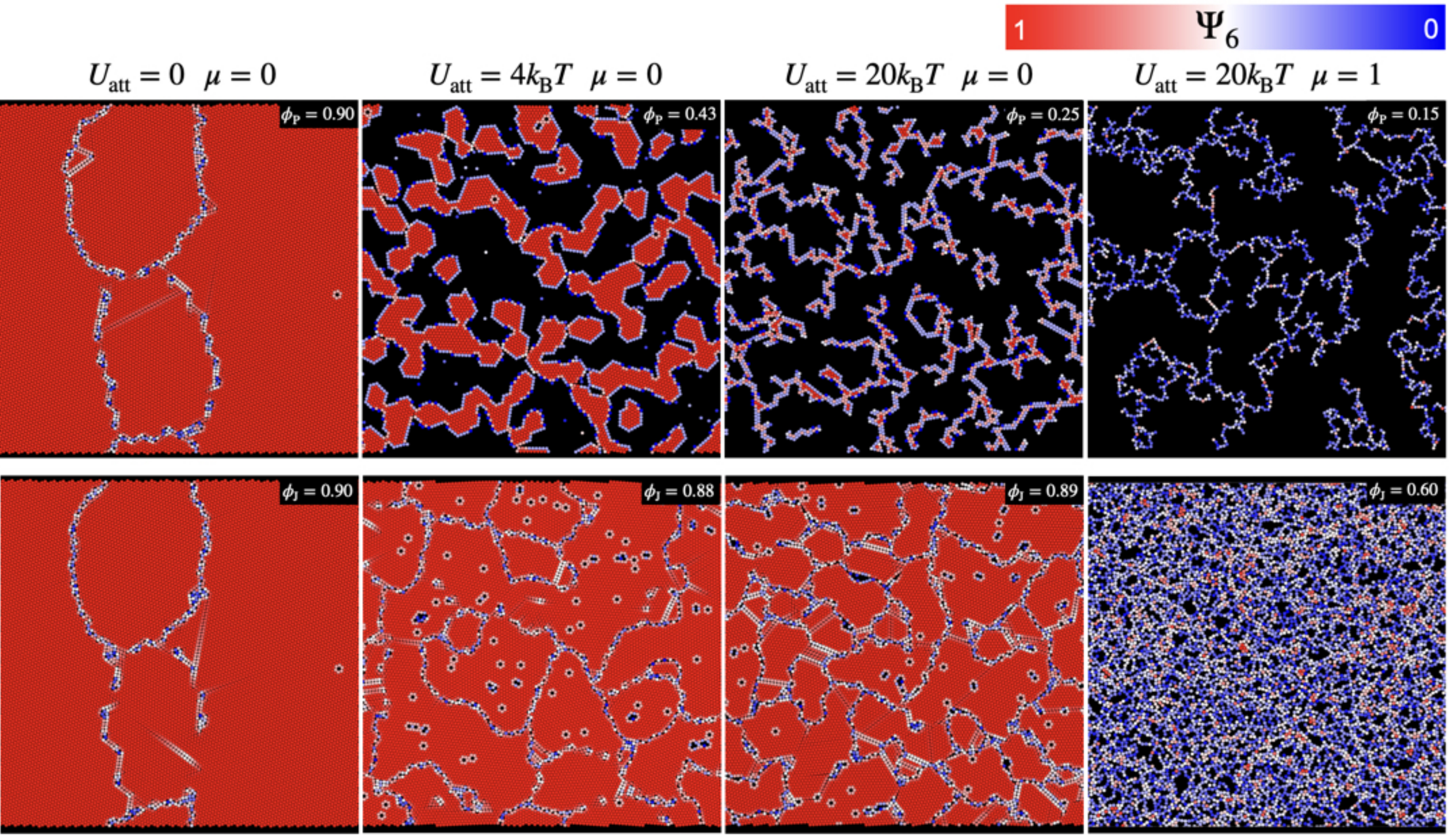}
\caption{The snapshots of different systems at percolation (upper panel) and jamming points (lower panel). The size of each square is $100d \times 100d$. Colorbar on the top indicates $\Psi_{6}$ of each particle.}
\label{fig:4}
\end{figure*}

Through the hexagonal order explained 
in Sec.~\ref{sec:micrography}, we visualize the microstructures, see \figref{fig:4} as well as Supplemental Videos S1--S4. The $\bar{\Psi}_{6}$ at jamming point $\phi_{\rm J}$ is shown in Fig.~S4. Non-attractive particles only exhibit high $\bar{\Psi}_{6}$ at the very end of compression (large $\phi$). For attractive cases $U_{\rm att} > 0$, the order parameters at the jamming $\phi_{\rm J}$, corresponding to $\mathsf{P}_1$ in experiments, decrease with both $U_{\rm att}$ and $\mu$, shown in Fig.~S4. This tendency is also observed also in our experiments (See Fig.~S2(a) of Supplementary Information). The less ordered structures at the jamming point in Fig.~S4 is consistent with our experimental observations in Fig.~S2 qualitatively.

\section{Further discussions}
In this section, we discuss the comparisons of our experimental and numerical results more
(Sec.~\ref{sec:results_and_discussions} 
and \ref{sec:simulation}). As we have discussed in Sec.~\ref{sec:simu_results_and_discussions}, the dependencies of attraction and friction to $\phi_\mathrm{P}$ and $\phi_\mathrm{J}$ have been reproduced qualitatively by our coarse-grained simulations. Though we have modeled the surface roughness by the cohesion and friction independently, they should be dependent because of the common origin, roughness. Recently, the real relationship between surface topography and effective friction has been measured using some special particles 
 experimentally \cite{Hsu2018} and studied numerically \cite{Papanikolaou2013}. Because friction can depend on the details of the surface, such friction measurements can be useful for our particles with irregular roughness. 
Also, how details of the attraction force determine the intermediate states of isotherms and the percolation points quantitatively can be interesting future works. However, it should be emphasized that the $\Pi$--$A$ isotherms cannot be used to extract the interacting potentials, as reported by Mears et al. \cite{Mears2020}. Therefore, a bottom-up approach can be a possible theoretical challenge, such as deriving the potential incorporating capillary multipole--multipole interactions and the surface pressure with an assumed configuration.

Besides, we remind that there are experimental difficulties in determining and controlling the exact area fraction both initially and at every step of compressions, which makes it impossible to explore jamming and percolation points more rigorously, unlike simulations. On the other hand, the exponent $\lambda$ for $\Pi \sim \phi^\lambda$ can be extracted without knowing the precise conversion of area fraction and area, which is examined and compared with the numerical $\Pi$--$\phi$ curves in Fig.\,S5 in Supplementary Information. The intermediate states of both our experiments and simulations give a power law with $\lambda \approx 5$ approximately, which is consistent with the typical values of colloidal gel compactions \cite{Buscall_1987a,Channell1997,Seto2013}.

\section{Conclusions}
We reveal the effect of surface roughness on the interaction between particles at a liquid interface by combining compression experiments and simulations. Experimentally, four roughnesses are systematically designed \cite{Chenwei_rough} and compared via $\Pi$--$A$ isotherm measurements and SEM micrographs. We find an intermediate state between a gas-like and a close-packed jamming state, corresponding to the percolating network structure \cite{Basavaraj2006}. This is considered to originate from the roughness-induced capillary attraction, which is supported by the estimation of the ratio of the capillary interaction energies according to the particle morphology (cf.~\cite{Stamou2000,Botto2012r}). As the compression proceeds, roughness-induced contact force plays an increasingly important role. We also find that surface roughness decreases both the jamming point $\phi_{\rm J}$ and the order parameter at $\phi_{\rm J}$ (see Fig.~S2 and S4). For rough particles, there are still voids at the jamming point. This happens because roughness enhances friction or interlocking effects \cite{Hsu2018,Papanikolaou2013,Hsiao2017,Singh2020,Ilhan2022}. In addition, experimental results also suggest that surface roughness enables particles to escape from the monolayer in a more gradual manner, possibly due to the tangential contact force.
Numerically, the features of roughness are characterized by attraction and friction between particles. Our simulation suggests that such interparticle attraction results in a percolating structure prior to the close-packed jamming, reminiscent of colloidal gels. Specifically, the percolation point $\phi_{\rm P}$ decreases by attraction. 
Similar to what we observed in our experiments, the decreases of both the jamming point $\phi_{\rm J}$ and the crystallization order $\bar{\Psi}_{6}$ happen by friction when the attraction also exists.

Thanks to the wide range of compaction, our study provides fundamental insights into different soft matter physics, such as colloidal gelation, frictional jamming, and buckling or collapse. 
The effects of the irregular roughness to gelation and friction in this work also remain a mystery compared with other types of rough particles, such as raspberry shapes \cite{Hsu2018,Weijgertze2020} which have been recently synthesized and intensively studied in 3D. We believe that protocol dependencies on $\phi_\mathrm{P},\,\phi_\mathrm{J}$ can be worth to be explored in terms of compression speeds, initial state preparations and compression geometry (uniaxial/biaxial), etc. Furthermore, the dynamics and mechanisms of collapses of monolayers are worth to be explored in relation to tangential contact forces surface morphology. Our study on rough colloids at liquid interfaces will also benefit the development of new biomaterials and functional devices.



\section*{Author Contributions}
T.L.~conceived the research. W.C.~synthesized the particles. A.N.K.~performed the experiments. Y.J.~conducted the simulations. A.N.K., Y.J., R.S.,~and T.L.~analyzed the results and wrote the manuscript.

\section*{Declaration of Competing Interest}
The authors declare that they have no known competing financial interests or personal relationships that could have appeared to influence the work reported in this paper.

\section*{Acknowledgements}
The authors thank Wenzhou Institute, UCAS for the access to shared experimental facilities. This work was financially supported by Wenzhou Institute, University of Chinese Academy of Sciences (WIUCASQD2020003, WIUCASQD2020002) and National Natural Science Foundation of China (11904390, 1217042129, 12150610463).





\input{output.bbl}




\end{document}



\title{Supporting Information for ``How Surface Roughness Affects the Interparticle Interactions at a Liquid Interface''}
\author{Airi N. Kato$^{1,3}$}
\email{These authors contributed equally to this work.}
\author{Yujie Jiang$^{2,3}$}%
\email{These authors contributed equally to this work.}
\author{Wei Chen$^{1,3}$}
\author{Ryohei Seto$^{2,3,4}$}%
\email{seto@wiucas.ac.cn}
\author{Tao Li$^{1,3}$}%
 \email{litao@ucas.ac.cn}
\affiliation{%
 $^{1}$Wenzhou Key Laboratory of Biophysics, Wenzhou Institute, University of Chinese Academy of Sciences, Wenzhou, Zhejiang 325001, China
}%
\affiliation{%
 $^{2}$Wenzhou Key Laboratory of Biomaterials and Engineering, Wenzhou Institute, University of Chinese Academy of Sciences, Wenzhou, Zhejiang 325001, China
}%
\affiliation{%
 $^{3}$Oujiang Laboratory (Zhejiang Lab for Regenerative Medicine, Vision and Brain Health), Wenzhou, Zhejiang 325001, China
}%
\affiliation{$^{4}$Graduate School of Information Science, University of Hyogo, Kobe, Hyogo 650-0047, Japan}

\date{\today}

\maketitle

\section{Experimental details}
\begin{figure}[htbp]
\includegraphics[width=0.7\textwidth]{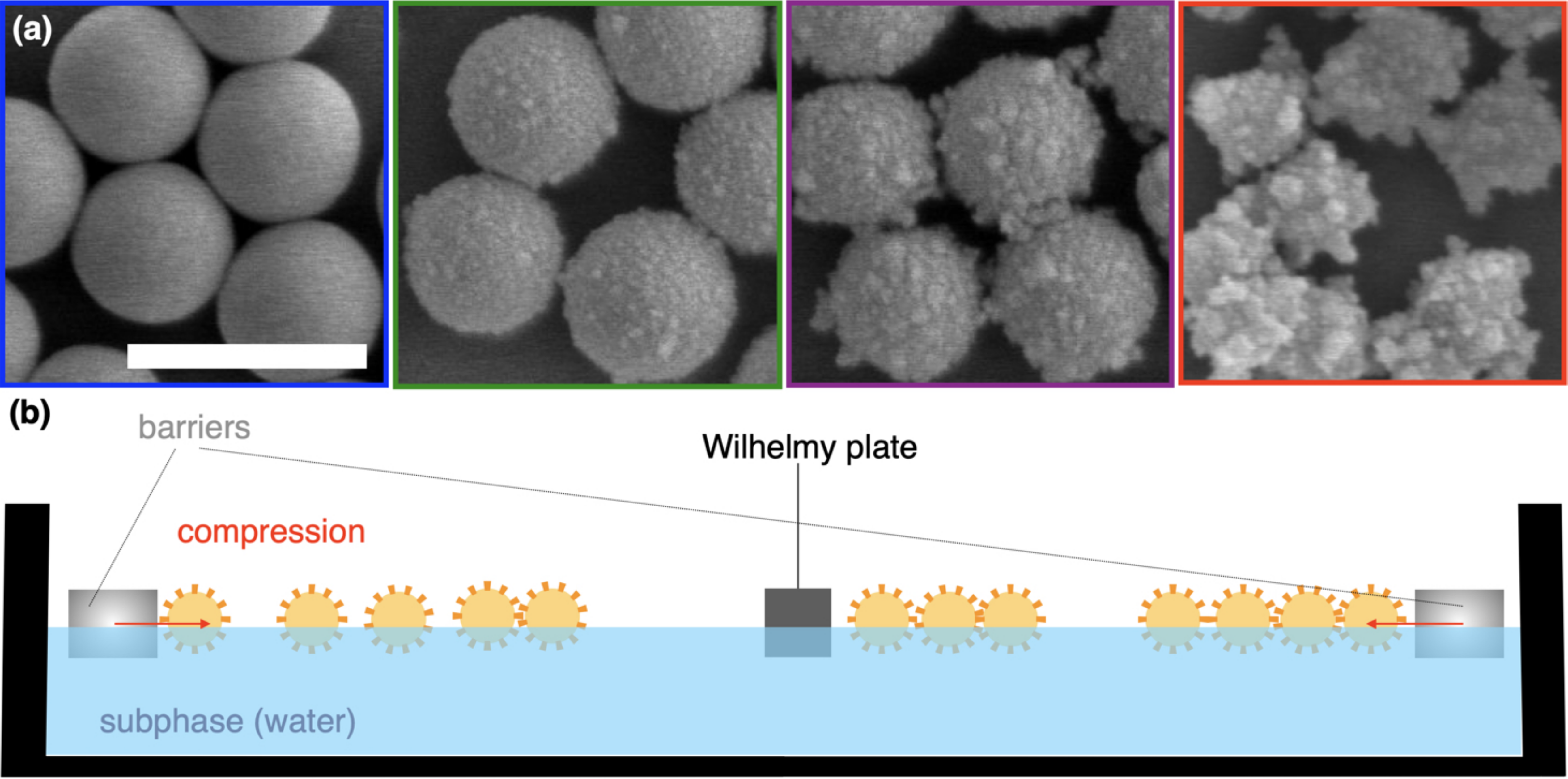}
\caption{(a) SEM images of SM, TR, MR, and VR particles in high magnification (from left to right). Scalebar=\SI{500}{\nano\metre}. (b) Schematics of a Langmuir trough}
\label{fig:setup}
\end{figure}

The schematic figure of the Langmuir trough is shown in \figref{fig:setup}(a).
To show our particles' surface clearer, the SEM images in a high magnification are shown in \figref{fig:setup}(b).

\section{Calculation of the hexagonal order parameter from the SEM micrographs}
\begin{figure}[htbp]
\includegraphics[width=0.99\textwidth]{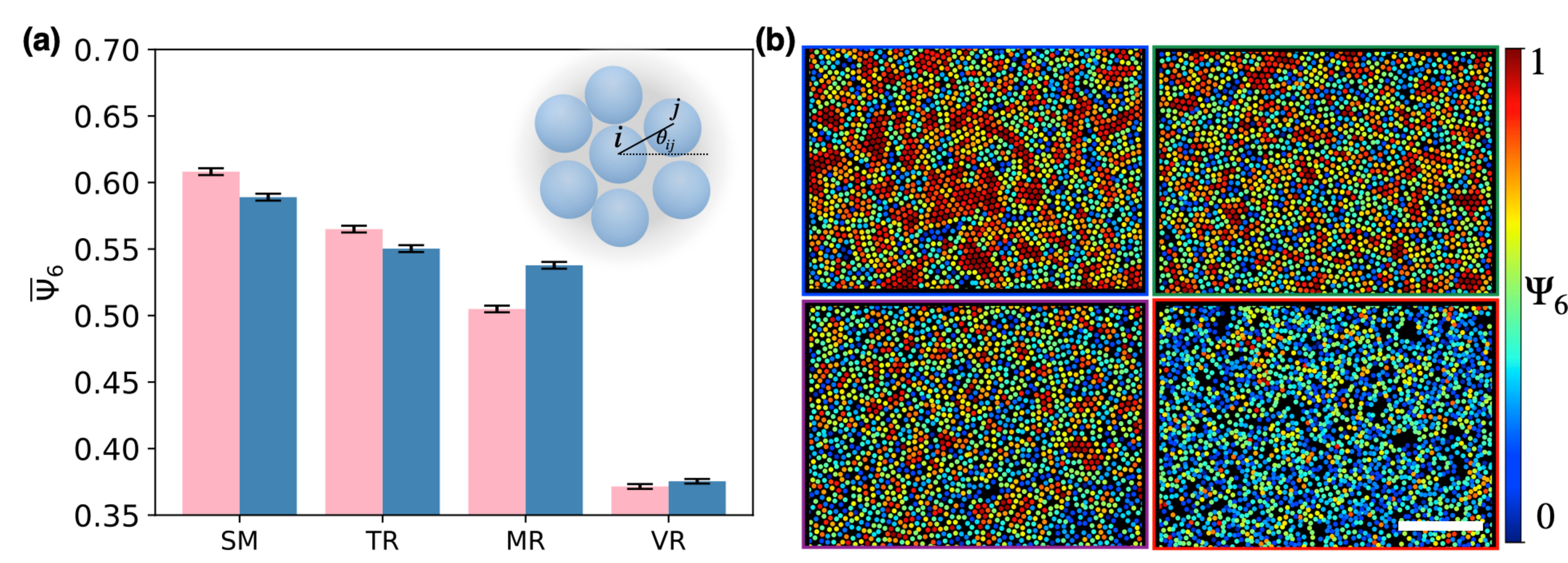}
\caption{(a) The global hexagonal order parameter $\bar{\Psi}_{6}$ at $P_1$ ($\sim \phi_\mathrm{J}$, pink) and around $P_2$ ($\sim$ collapse, blue) in the experiments. The error bars mean the standard errors. The schematic figure of the inset is for the definitions of $\bar{\Psi}_{6}$. ($N>7300$ for each calculation.) (b) The local hexagonal order parameter $\Psi_6$ at the close-packed jamming $P_1$ (rendered). The frame color of the four panels corresponds to the colors of isotherms (Fig.1(b) of the main text). Colorbar indicates $\Psi_6$ of each particle. Scale bar=$5\mu{\rm m}$.}
\label{fig:hex}
\end{figure}

The positions of particles were decided by binarizations of the SEM images using Fiji \cite{Fiji}. The global hexagonal order parameter $\bar{\Psi}_{6}$ is calculated at the close-packed jamming points $P_1$ and around the onsets of collapses of monolayers $P_2$, as shown in \figref{fig:hex}(a). We used a python package freud \cite{freud2020} for the calculation. Note that the collapses had already started except for MR. Also, the color-coded rendered images of particles' configuration at $P_1$ are shown in \figref{fig:hex}(b).

As expected from the SEM observations, $\bar{\Psi}_{6}$ at the close-packed jamming $P_1$ decreased with particles' surface roughness (pink bars, \figref{fig:hex}(a)). We can also see the more ordered regions for the smoother particles in \figref{fig:hex}(b). Compared with our simulation results in Fig.4~(a) of the main text, the experimental structures are less ordered, which can be due to polydispersity and heterogeneity of the effective friction between particles.  
Further compression induces abrupt collapse in the monolayers of SM and TR particles, which  decreases $\bar{\Psi}_{6}$ (see the corresponding blue bars). 
The MR particles preferentially occupy the voids regions in plane after $P_1^{\mathrm{MR}}$, which makes denser and more ordered monolayer (the change between Fig.~2(c) and (f) in the main text). This leads to the increase of $\bar{\Psi}_{6}$ for the MR particles.
In contrast, for the VR particles, the order parameter $\bar{\Psi}_{6}$ remains nearly unchanged after close-packed jamming (\figref{fig:hex}). 

\section{Simulation Methods}

\subsection*{Model}

We simulate a colloidal monolayer composed of $N = 12478$ particles (disks) of diameter $d = 1$ (radius $R = 0.5$) in a rectangle box with periodic boundaries. We use the combination of Langevin thermostat and the microcanonical ensemble (NVE) to fix the system at $k_{\rm B}T = 1$. We set particle mass $m = 1$ and solvent viscosity $\eta = 8.2$, which gives a Brownian timescale $t_{\rm B} = \pi\eta d^3/2k_{\rm B}T$. The particle interaction is characterized by the Derjaguin--Muller--Toporov (DMT) contact model \cite{derjaguin1975effect}. The normal component of the DMT model is expressed as:
\begin{equation}
\mathbf{F}_n = \left(\frac{3}{4}ER^{1/2}\delta^{3/2}-4\pi\gamma R \right)\mathbf{n},
\end{equation}
where $\delta$ refers to the particle overlap, $E$ the Young's modulus and $\gamma$ the surface energy density. We set $E = 14802147.7$ to ensure hard particles and vary $\gamma$ from 0 to 1636, corresponding to adhesion energy $U_{\rm adh} = 0k_{\rm B}T$ to $20k_{\rm B}T$. The tangential component of the DMT model consists of sliding and rolling friction, both characterized by a spring coefficient $k$ and a friction coefficient $\mu$ as follows:
\begin{equation}
\mathbf{F}_t = -\min{\left(\mu F_n,-k\xi\right)}\mathbf{t},
\end{equation}
where $\mathbf{F}_t$ represents either frictional force (sliding) or torque (rolling) and $\xi$ refers to the corresponding displacement. We set $k = 10^5$ to be sufficiently large and vary $\mu$ from 0 to $1$ to mimic particles of different roughness.

Prior to compression, we first create a random configuration and allow particles to relax as Brownian hard disks ($\gamma = 0$) until reaching a homogeneous, non-overlapping, random state. We then let the system to equilibrate until reaching a state with steady pressure. Uniaxial compression is then performed by reducing the length $L_x$ in $x$ direction at a constant velocity $v = 0.005$ while keeping the box width in $y$ direction $L_{y} = 100 \gg d$ constant. That is, the system is condensed from $\phi = N\pi d^2/4L_xL_y = 0.14$ to $\phi = 1.0$. Area fractions $\phi$ beyond $\phi_{\rm cp}$ represents compression of particles (non-zero overlap).

\subsection*{Parameters}

\begin{table}[htb]

\centering
\begin{tabular}{ c | c | c | c }
\centering
Quantity & Symbol & Dimension & Value \\
\hline \hline
Particle diameter       & $d$           & [L]               & 1.0 \\
\hline
Particle mass           & $m$           & [M]               & 1.0 \\
\hline
Particle number         & $N$           &                   & 12478 \\
\hline
Thermal energy          & $k_{\rm B}T$  & [M$\rm L^2T^{-2}$]& 1.0 \\
\hline
Box size                & $L_{x,y}$     & [L]               & 700$\to$100, 100 \\
\hline
Viscosity               & $\eta$        & [M$\rm LT^{-1}$]  & 8.2 \\
\hline
Young's modulus         & $E$           & [M$\rm LT^{-2}$]  & 14802147.7 \\
\hline
Surface energy density  & $\gamma$      & [M$\rm T^{-2}$]   & 1636 \\
\hline
Stiffness of tangential constrain  & $k$      & [M$\rm T^{-2}$]   & 100000 \\
\hline
Compression rate        & $v$           & [T]$^{-1}$          & 0.01 \\
\hline
Time step               & $\tau$        & [T]               & 0.00026 \\
\end{tabular}
\caption{Simulation parameters.}
\label{tab:parameters}
\end{table}

All parameters involved in our simulation are summarized in \tabref{tab:parameters}. Langevin dynamics simulation is performed via time integration with a time step $\tau = 2.6 \times 10^{-4}$, which is sufficiently small compared with both the collision timescale $t_c = \sqrt{m/Ed}$ and damping timescale $t_{\rm damp} = m/3\pi\eta d$. This ensures stable simulation without losing particles. The system is sufficiently damped since the damping time is three orders of magnitude lower than the Brownian time, i.e. $t_{\rm damp} \lesssim 10^{-3}t_{\rm B}$. Meanwhile, we choose a low compression rate ($L_x/v \gg t_{\rm B}$) to make sure that the compression process is quasi-static.

\subsection*{Dynamics}
 
The dynamics of each particle is described by the Langevin equation:
\begin{equation}
\mathbf{m}\cdot\frac{\dif \mathbf{U}}{\dif t} = \mathbf{F}_{\rm H} + \mathbf{F}_{\rm B} + \mathbf{F}_{\rm P},
\end{equation}
where $\mathbf{m}$ is the mass (tensor) and $\mathbf{U}$ the particle velocity. Right-side terms respectively refer to hydrodynamic drag force $\mathbf{F}_{\rm H} = -6\pi\eta R\left[\mathbf{U^\infty} - \mathbf{U}\right]$, Brownian force $\mathbf{F}_{\rm B}$ which enforces $k_{\rm B}T = 1$ and interparticle force $\mathbf{F}_{\rm P}$ from the DMT model. All simulations are performed with the open source program LAMMPS, in which the equations of motion above are integrated via Velocity-Verlet algorithm \cite{LAMMPS}.

\subsection*{Isotherm}

Isotherm monitors the evolution of surface pressure $\Pi$ which consists of osmotic pressure $\Pi_{\rm O}$ and particle pressure $\Pi_{\rm P}$:
\begin{equation}
\begin{aligned}
\Pi &= \Pi_{\rm O} + \Pi_{\rm P},\\
\Pi_{\rm O} &= \frac{Nk_{\rm B}T}{A} = \frac{1}{A}\left\langle\sum_i^N m_i{\bf v}_i^2\right\rangle,\\
\Pi_{\rm P} &= \frac{1}{A}\left\langle\sum_i^N{\bf r}\cdot{\bf F}\right\rangle,
\end{aligned}
\end{equation}
where $\left\langle\dots\right\rangle$ represents the time average. For each presented $\Pi$--$A$ isotherm data, we measure the average of $\Pi$ over $t = d/v \approx 5t_{\rm B}$, which corresponds to the time for $x$ boundary to move one diameter of the particle.

\subsection*{Determination of percolation and jamming points}

We use particle configuration to determine whether the system is percolated or jammed. Particles with distance $r < d = 1$ (i.e., overlap $\delta > 0$) are grouped as one cluster, and we define percolation as the point when there exists at least one cluster that connects through the periodic boundary in the $x$ direction (since we compress $x$ direction). We reckon that jamming occurs when compression starts to be exerted on each single particle. In this way, we use the evolution of average overlap $\langle \delta \rangle$ between particles to determine jamming---the point where $\langle \delta \rangle$ shows a drastic increase. This point, in particular, can be well defined by the peak position of the second derivative of $\langle \delta \rangle$ by $A$. 

\begin{figure}[htbp]
\includegraphics[width=0.6\textwidth]{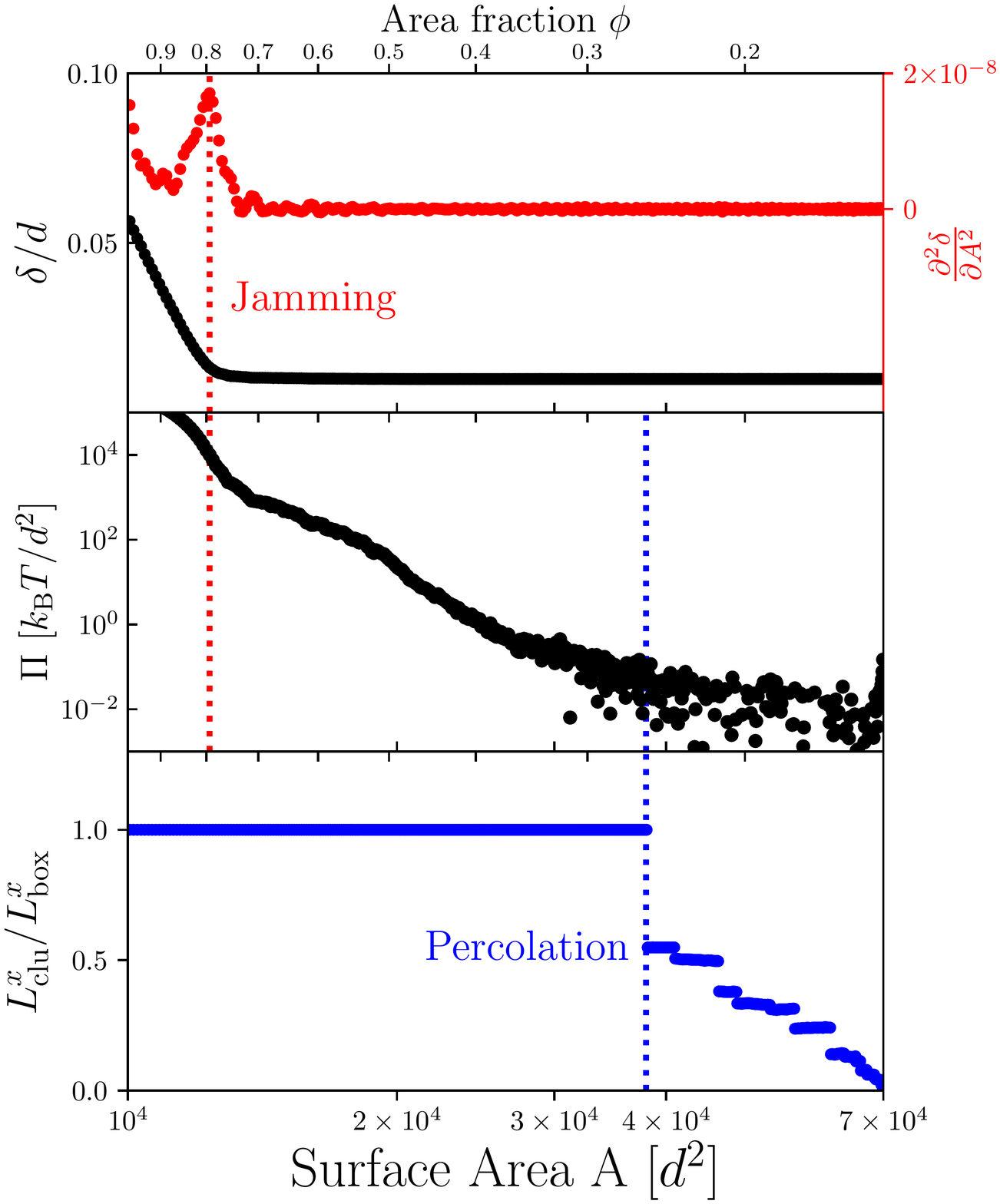}
\caption{Demonstration of our definition on jamming $\phi_{\rm J}$ and percolation points $\phi_{\rm perc}$. The two critical points demarcate the intermediate state of $\Pi$--$A$ isotherm.}
\label{fig:s1}
\end{figure}

Below we use adhesive particles of $U_{\rm adh} = 20k_{\rm B}T$ and $\mu = 1$ to demonstrate the two critical points, \figref{fig:s1}. As compression proceeds, the particles, randomly distributed initially, aggregate into clusters which grow over time. We use the evolution of the length in $x$ direction of the largest cluster $L^x_{\rm clu}$ to represent the progress of percolation, and regard the moment when $L^x_{\rm clu} = L_x$ as the critical percolation point, dashed blue line in \figref{fig:s1}. The jamming point is labeled by the peak position of $\delta^{\prime\prime}(A)$, red dashed line in \figref{fig:s1}. The two critical points, derived purely from particle configuration, demarcate the boundaries of intermediate state which is consistent with the $\Pi$--$A$ isotherm.

\section{The hexagonal order parameters at jamming $\phi_\mathrm{J}$ in simulation}
\begin{figure}[htbp]
\includegraphics[width=0.5\textwidth]{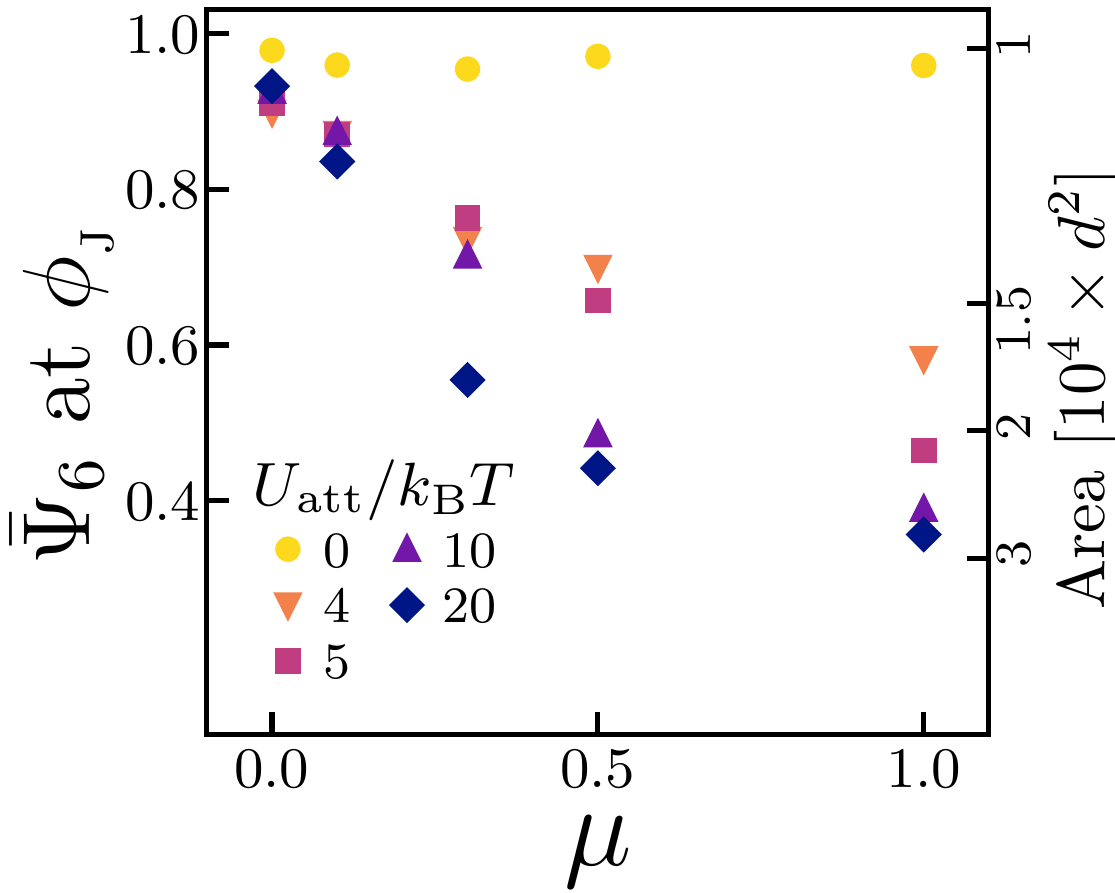}
\caption{The hexagonal order parameters at jamming $\phi_\mathrm{J}$}
\label{fig:s4}
\end{figure}
The global order parameters $\bar{\Psi}_{6}$ at each jamming point $\phi_{\rm J}$ are shown in \figref{fig:s4}. With attraction, the order decreases with the friction coefficient $\mu$, similar to experimental results \figref{fig:hex} qualitatively.




\section{The power law in the intermediate state}

\begin{figure}[htbp]
\includegraphics[width=0.99\textwidth]{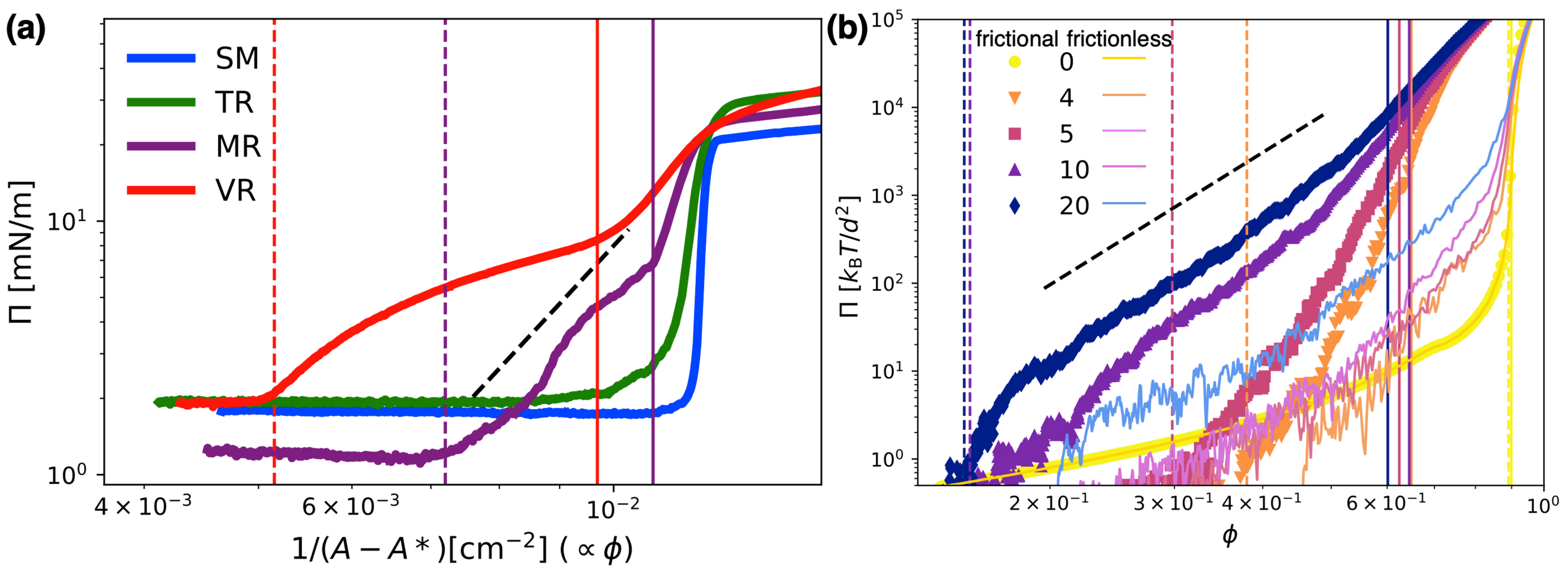}
\caption{Power laws in the intermediate state. (a) experiments, (b) simulations (frictionless and frictional, the unit of the legends is $k_{\rm{B}}T$.) In (a), the values $1/(A-A^{\ast})$ proportional to $\phi$ are used to check only the exponent. $A^{\ast}$ is set arbitrary for the visibility, which do not affect the exponent. The black dashed lines of both (a) and (b) are guides for eyes with the slope 5 (i.e., $\Pi\propto \phi^5$). The dashed and solid vertical lines with colors corresponding to the percolation and jamming points of isotherms with the correspondent colors. (In the panel (b), the percolation and jamming points only for the frictional cases are shown for better visibility.)}
\label{fig:s2}
\end{figure}

We identified that the intermediate state is a gel state, forming a percolated network. Gel states typically give the power law:  $\Pi(\phi)\sim \phi^{\lambda}$ \cite{Channell1997,Botet2004,Seto2013} ($\Pi(\phi)$ can be a yield stress $\Pi_{\rm y}(\phi)$). Thus, the log-log plots of isotherms as a function of the area fraction $\phi$ are shown in \figref{fig:s2} (a,~b): experiments and simulations. The intermediate state of MR particles in the experiment and strong attractive cases in simulation show nearly a power law with $\lambda \approx 5$.

\section{Supplemental Video Legend}

{\bf Supplemental Videos S1--S4} Time-lapsed videos of four different systems during compression in simulations. Particles are colored according to their local order parameter $\Psi_6$, see color bar on the top of each movie.
\begin{itemize}
    \item [{\bf S1:}] Non-attractive frictionless particles, with $U_{\rm att} = 0$ and $\mu = 0$.
    \item [{\bf S2:}] Attractive frictionless particles, with $U_{\rm att} = 20k_{\rm B}T$ and $\mu = 0$.
    \item [{\bf S3:}] Non-attractive frictional particles, with $U_{\rm att} = 0$ and $\mu = 1$.
    \item [{\bf S4:}] Attractive frictional particles, with $U_{\rm att} = 20k_{\rm B}T$ and $\mu = 1$.
\end{itemize}

\input{output_SI.bbl}

%% file: output_SI.bbl
%